\documentclass{article}

\usepackage{PRIMEarxiv}

\usepackage[utf8]{inputenc} 
\usepackage[T1]{fontenc}    
\usepackage{hyperref}       
\usepackage{url}            
\usepackage{booktabs}       
\usepackage{amsfonts}       
\usepackage{nicefrac}       
\usepackage{microtype}      
\usepackage{lipsum}
\usepackage{fancyhdr}       
\usepackage{graphicx}       
\usepackage{multirow}

\usepackage{amssymb}
\usepackage{amsmath}
\usepackage{stfloats}
\usepackage{xcolor}
\usepackage{algorithm}  
\usepackage{algorithmic}
\usepackage[normalem]{ulem}
\usepackage{lscape}
\usepackage{geometry}
\usepackage{caption} 
\usepackage{longtable}
\usepackage{setspace} 
\usepackage[justification=centering]{caption}

 \usepackage{lineno}
 
\usepackage{subcaption}
\usepackage{multirow}
\usepackage{siunitx}
\usepackage{glossaries}

\usepackage{pifont}

\newcommand{\setglscolor}{\def\glstextformat{\textcolor{black}}}

\setglscolor

\newcommand{\RequestCenter}{\textsc{Request Center}}
\newcommand{\ChatGPT}{\textsc{ChatGPT}}
\newcommand{\VITR}{\textsc{VITR}}
\newcommand{\langid}{\textsc{langid}}

\newacronym{VSE}{VSE}{Visual-Semantic Embedding}
\newacronym{VSRN++}{VSRN++}{Visual-Semantic Reasoning Network}
\newacronym{VSEinf}{VSE$\infty$}{Variation of Visual-Semantic Embedding Network}
\newacronym{CLIP}{CLIP}{Contrastive Language-Image Pre-training network}
\newacronym{VITR}{VITR}{VIsion Transformers with Relation-focused learning network}

\graphicspath{{media/}}     

\title{Boon: A Neural Search Engine for Cross-Modal Information Retrieval
}

  \author{
  Yan Gong, Georgina Cosma \\
  Department of Computer Science \\
  Loughborough University \\
  Loughborough\\
  \texttt{\{y.gong2, g.cosma\}@lboro.ac.uk}}
  
\begin{document}
\maketitle

\begin{abstract}
Visual-Semantic Embedding (VSE) networks can help search engines better understand the meaning behind visual content and associate it with relevant textual information, leading to more accurate search results. VSE networks can be used in cross-modal search engines to embed image and textual descriptions in a shared space, enabling image-to-text and text-to-image retrieval tasks. However, the full potential of VSE networks for search engines has yet to be fully explored. This paper presents Boon, a novel cross-modal search engine that combines two state-of-the-art networks: the GPT-3.5-turbo large language model, and the VSE network VITR (VIsion Transformers with Relation-focused learning) to enhance the engine's capabilities in extracting and reasoning with regional relationships in images. VITR employs encoders from CLIP that were trained with 400 million image-description pairs and it was fine-turned on the RefCOCOg dataset. Boon's neural-based components serve as its main functionalities: 1) a `cross-modal search engine' that enables end-users to perform image-to-text and text-to-image retrieval. 2) a `multi-lingual conversational AI' component that enables the end-user to converse about one or more images selected by the end-user. Such a feature makes the search engine accessible to a wide audience, including those with visual impairments. 3) Boon is multi-lingual and can take queries and handle conversations about images in multiple languages. Boon was implemented using the Django and PyTorch frameworks. The interface and capabilities of the Boon search engine are demonstrated using the RefCOCOg dataset, and the engine's ability to search for multimedia through the web is facilitated by Google's API.

\end{abstract}


\section{Introduction}

\begin{figure}
\centering 
\includegraphics[width=0.6\linewidth]{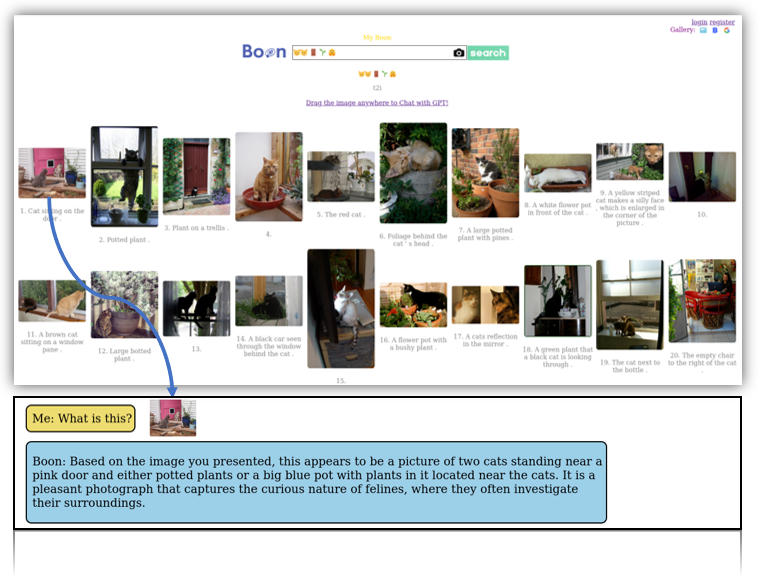} 
\caption{The proposed search engine, Boon, enables cross-modal information retrieval and facilitates conversations about images with users.} 
\label{Boon} 
\end{figure}

Search engines have transformed the way people discover and access multimedia resources (such as texts, images, and videos) by providing fast and easy search capabilities \cite{messina2021fine, gao2020fashionbert}. Traditional search engines typically rely on textual information such as metadata, tags, to identify and retrieve relevant images \cite{zhang2013image, lowe1999object}. 
Cross-modal information retrieval-based search engines enhance multimedia search experiences by leveraging advanced techniques like Natural Language Processing (NLP) and Computer Vision (CV) to bridge the gap between text and image modalities, allowing users to obtain more relevant and accurate results \cite{karpathy2015deep, vinyals2015show}. 
State-of-the-art neural networks for cross-modal information retrieval are \gls{VSE} networks, which embed image-description pairs in a shared latent space and compute similarity scores for image-to-text and text-to-image retrieval tasks \cite{gong2021limitations}.
Li et al.\cite{li2022image} proposed \gls{VSRN++}, which uses a Graph Convolutional Network (GCN) \cite{kipf2017semi} to extract the relationships between image objects, resulting in high-level visual semantics. 
Chen et al. \cite{chen2021learning} presented a \gls{VSEinf}, which utilises a generalised pooling operator to uncover the optimal strategy for combining image and description representations.  
Radford et al. \cite{radford2021learning} proposed \gls{CLIP}, which enables efficient learning of visual concepts through natural language supervision using 400 million image-description pairs.  
Recently, Gong et al. \cite{gong2023vitr} introduced \gls{VITR} that enhances Vision Transformers (ViTs) by employing a local encoder to extract and reason about image region relations, combining reasoned results with pre-trained global knowledge (e.g. from \gls{CLIP}) to predict similarity scores between images and descriptions. 
\gls{VITR} outperformed various state-of-the-art networks, including \gls{CLIP}, \gls{VSEinf}, and \gls{VSRN++}, in cross-modal information retrieval tasks, particularly in relation-focused cross-modal information retrieval \cite{gong2023vitr}. 

As a result, this paper focuses on developing a search engine that incorporates \gls{VITR}, improving user experience by emphasising information retrieval based on relations expressed in user queries and enhancing image-to-text and text-to-image retrieval performance.

Additionally, recently developed Large Language Models (LLMs), such as ChatGPT \cite{team2022chatgpt}, have exhibited exceptional capabilities in natural language understanding and generation \cite{shen2023chatgpt}, revolutionising various applications from conversational AI and content creation to sentiment analysis \cite{king2023conversation}. By integrating an LLM into a cross-modal information retrieval search engine, the engine can translate and summarise textual queries, addressing the constraints of existing \gls{VSE} networks that support only brief English queries. On the other hand, LLMs, such as ChatGPT's 3.5 model, exhibit limitations in comprehending image modalities. However, these shortcomings can be mitigated by employing \gls{VSE} networks to obtain the most relevant description for an image and utilising this description as a textual prompt for the LLM.

Therefore, this paper presents Boon (shown in Figure \ref{Boon}), a novel cross-modal search engine that combines two state-of-the-art networks: ChatGPT (an LLM), and \gls{VITR} (a \gls{VSE} network) to enhance the engine's capabilities in extracting and reasoning with regional relationships in images. 
The contributions of this paper are as follows:

\begin{itemize} 
    
    \item The proposed Boon is a search engine that benefits from high cross-modal information retrieval performance due to its integration of \gls{VITR}.   
    It enables users to retrieve images using textual queries or to retrieve textual descriptions and their corresponding images using image queries from a gallery. 
    Additionally, Boon re-ranks the results of Google's Programmable Search Engine API (Google's API) to make them more relevant to the query, and this improves search results, particularly for queries that contain relation-related content. 
    
    \item The proposed search engine uses ChatGPT to support textual queries written in multiple languages. One of \gls{VITR}'s limitations is that it can only support textual queries in English. By combining the capabilities of \gls{VITR} and ChatGPT, non-English queries detected using the Python \langid\, library can be translated into English. 
    
    \item ChatGPT's 3.5 model has limitations in its ability to comprehend image modalities. Boon can converse with end-users about images and this feature can ultimately enhance their experience while using the engine.
\end{itemize} 

\section{Related Work}

This section discusses related work on cross-modal information retrieval networks and large language models.

\subsection{Cross-modal Information Retrieval Networks}

Current works use \gls{VSE} networks to embed image-description pairs in a shared latent space and calculate similarity scores for retrieval tasks \cite{gong2021limitations, faghri2018vse++, zhang2021aggregation, wang2020pfan++, zhang2020context, wei2020multi}.
Faghri et al. \cite{faghri2018vse++} proposed an enhanced \gls{VSE} architecture which employs a fully connected neural network and a Gated Recurrent Units (GRU) network \cite{cho2014learning} to embed image features (extracted by the Faster R-CNN \cite{ren2016faster, anderson2018bottom}) and descriptions as representations, respectively.
Lee et al. \cite{lee2018stacked} explored the full latent alignments between image regions and descriptive words to determine the similarity of image-description pairs. 
Li et al. \cite{li2019visual, li2022image} enhanced image features with image region relations extracted by a GCN \cite{kipf2017semi}. 
Chen et al. \cite{chen2021learning} proposed a variation of the \gls{VSE} network that benefits from a generalised pooling operator, which uncovers the best strategy for pooling image and description representations.

The development of pre-trained networks for cross-modal information retrieval has advanced significantly in recent years \cite{chen2020uniter, yu2021ernie, cheng2022vista, gan2020large, lu2022cots}. Chen et al. \cite{chen2020uniter} introduced a novel network that is an universal image-text representation learned through large-scale pre-training on four image-text datasets. 
Lu et al. \cite{lu2022cots} presented a novel collaborative two-stream vision-language pre-training approach for image-text retrieval that strengthens cross-modal interaction through instance-level alignment, token-level interaction, and task-level interaction.
Radford et al. \cite{radford2021learning} proposed the pre-trained \gls{CLIP}, which applies contrastive learning to align global visual representations and textual representations from a dataset containing 400 million image-description pairs.
Recently, Gong et al. \cite{gong2023vitr} proposed \gls{VITR}, a novel network that augments the ViT by extracting and reasoning about image region relations. \gls{VITR} addresses the limitations of ViT-based networks in relation-focused cross-modal information retrieval tasks \cite{mao2021dual, zhong2022regionclip} and outperforms state-of-the-art networks such as \gls{CLIP} in these tasks.

\begin{figure*}[ht] 
\centering 
\includegraphics[width=1\linewidth]{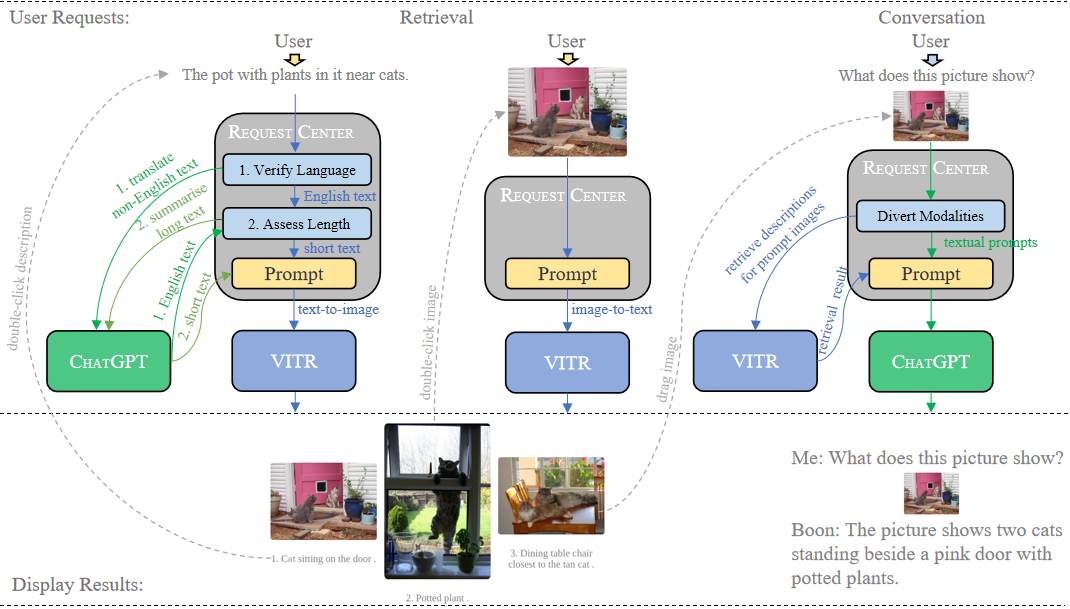} 
\caption{Flowchart of Boon's functionalities. For retrieval requests: 1) in text-to-image retrieval, users input a textual query to retrieve relevant images; and 2) in image-to-text retrieval, the image query is used to retrieve relevant descriptions and their corresponding images. For conversation requests, users input a textual query to converse with \ChatGPT\, and can also upload images to enrich the discussion.} 
\label{BoonFlow} 
\end{figure*}

\subsection{Large Language Models}
LLMs have witnessed remarkable advancements in recent years, exhibiting exceptional performance across a wide range of NLP tasks \cite{devlin2019bert, kasneci2023chatgpt}. Early models, such as Word2Vec \cite{church2017word2vec} and GloVe \cite{pennington2014glove}, paved the way by generating dense vector representations of words, while Recurrent Neural Networks (RNNs) \cite{medsker2001recurrent} and Long Short-Term Memory (LSTM) \cite{hochreiter1997long} networks enabled sequential data processing. The advent of attention mechanisms and transformers, introduced by Vaswani et al. \cite{vaswani2017attention} further revolutionised the field of NLP. Building on these breakthroughs, more recent models such as BERT (Bidirectional Encoder Representations from Transformers) \cite{devlin2019bert}, OpenAI's GPT-3 \cite{brown2020language} and ChatGPT \cite{team2022chatgpt} have harnessed the power of unsupervised pre-training and fine-tuning to achieve impressive performance in various tasks, including natural language understanding and generation. Although these large-scale models have demonstrated unprecedented capabilities in NLP tasks, they have not been extensively employed in cross-modal search engines.

\section{The Proposed Boon Cross-Modal Search Engine}

This section presents the architecture of Boon along with a discussion of its retrieval and conversation modules, and a presentation of its front-end features and capabilities.

\subsection{The Architecture of Boon}
The proposed search engine, Boon, is illustrated in Figure \ref{BoonFlow}. Boon caters to users' requests for multi-lingual cross-modal retrieval and conversation: 1) for retrieval, users can either input a textual query to search for relevant images or upload an image query to search for relevant descriptions and their corresponding images; and 2) for conversation, users can input textual prompts and upload prompt images to have a conversation about the image(s) with \ChatGPT. Boon comprises three modules which are \VITR, \ChatGPT, and \RequestCenter, and they function as follows.

The \textbf{\VITR} module is ultilised for cross-modal information retrieval, and it aims to embed image-description pairs into a shared latent space, enabling the prediction of the pairs' similarity scores for the purpose of retrieval ranking \cite{gong2023vitr}. \VITR\, consists of:
1) a text encoder that encodes a description into a global representation and a set of local representations;
and 2) a ViT encoder and a CNN-based local encoder encode an image and its regions into a global representation and a set of local representations, respectively. 
\VITR\, can utilise the encoders from \gls{CLIP} to obtain global and local representations of images and texts, which were trained on 400 million image-description pairs. \VITR\, was fine-tuned on the RefCOCOg dataset \cite{mao2016generation} to learn reasoning relations and aggregate reasoned results from local representations, along with global knowledge to enhance relation-focused cross-modal information retrieval performance.

The \textbf{\ChatGPT} module utilises the GPT-3.5-turbo model through the ChatGPT API for translation and summarisation of textual queries, and to generate sentences for conversations with users based on various prompts.

The \textbf{\RequestCenter} module activates user requests to generate prompts for the \VITR\, and \ChatGPT\, modules. The details of how the \RequestCenter\, activates different user requests will be introduced in sections \ref{Retrieval Requests} and \ref{Conversation Requests}.

\subsection{Retrieval Requests} \label{Retrieval Requests}
\textbf{Text-to-Image Retrieval.}
For text-to-image retrieval requests, users can input a textual query to retrieve relevant images. The back-end \RequestCenter\, processes the textual query as follows:
1) Verifies the language of the textual query using the Python \langid\, library \cite{lui2012langid}. If the query is not written in English, the \RequestCenter\, asks \ChatGPT\, to translate it into English with the prompt: `Translate the following text to English, provide the result directly without explanations: (the textual query).'  The prompt for \ChatGPT\, does not require the query to be in a specific language.
2) Assesses the length of the textual query. If a query exceeds the 77-token limit permitted by \VITR, the \RequestCenter\, requests \ChatGPT\, to summarise the query in order to meet the length requirement. 

The processed textual query serves as the prompt for \VITR, and then \VITR\, returns the text-to-image retrieval results to be displayed.

\textbf{Image-to-Text Retrieval.}
Image-to-text retrieval requests use image queries to retrieve relevant descriptions, which are then displayed along with their corresponding images. The back-end \RequestCenter\, servers the image query uploaded by users as the prompt for \VITR. \VITR\, ranks the descriptions based on their relevance to the image query, and the descriptions and their corresponding images are displayed. Users can find interest in the displayed images.

\textbf{Switch Modes to Access Various Galleries.}
Users can switch between the `My Album', `My Boon', and `My Google' modes to access images from various galleries. Each mode accesses images from an independent gallery: 
1) `My Album' mode provides cross-modal information retrieval performance for managing users' pictures, and users can create an account and establish their personal gallery by uploading up to a fixed number of images (e.g., \num{500}) themselves. Each user can search for and retrieve images that reside within their personal gallery. 
2) `My Boon' mode, creates a common gallery and for demonstration purposes it was populated using the RefCOCOg dataset \cite{mao2016generation}, which contains \num{25799} real-world images, with \num{21899} of them having corresponding relevant descriptions. This allows all users to search for and retrieve images. 
3) `My Google' mode, enables users to search for and retrieve the relevant images related to their textual query from the web, using Google's API. Boon re-ranks the results returned by Google's API to provide a more accurate image retrieval service for users.

\subsection{Database}
\begin{figure}[htbp] 
\centering 
\includegraphics[width=0.5\linewidth]{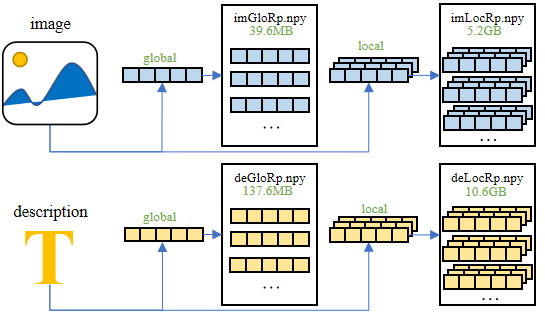} 
\caption{The encoded global and local representations for both images and descriptions generated by \VITR\, have been stored in the `.npy' files to improve retrieval speed.} 
\label{VITRdatabase} 
\end{figure}

The database is built for the common gallery of `My Boon'.
To optimise retrieval time, the representations for images and descriptions needed by \VITR\, have been pre-encoded and stored in the database. As illustrated in Figure \ref{VITRdatabase}, the database files `imGloRp.npy' (39.6MB), `imLocRp.npy' (5.2GB), `deGloRp.npy' (137.6MB), and `deLocRp.npy' (10.6GB) store the global representations of images, the local representations of images, the global representations of descriptions, and the local representations of descriptions, respectively. By directly accessing the saved representation values from the database files, Boon eliminates the need for encoding images and descriptions during the retrieval process, resulting in faster retrieval.

\subsection{Conversation Requests} \label{Conversation Requests}
For conversation requests, users can input a textual prompt to engage in a conversation with \ChatGPT, as well as upload images (multiple images are supported) for discussion. The back-end \RequestCenter\, initially checks whether the user's query includes images. If it does, the \RequestCenter\, prompts \VITR\, to retrieve the most relevant descriptions from a created description pool for the uploaded images, and the retrieved descriptions will serve as prompts for \ChatGPT. The description pool is derived from the MS-COCO dataset \cite{lin2014microsoft}, which encompasses \num{634083} diverse descriptions capable of accurately depicting real-world images.

\begin{figure}[htbp] 
\centering 
\includegraphics[width=0.5\linewidth]{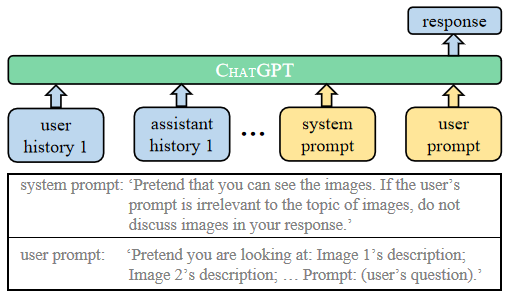} 
\caption{The method of facilitating a conversation about images with \ChatGPT\, using the prompts of the roles of the user, assistant, and system.} 
\label{ChatGPT} 
\end{figure}

The prompt for \ChatGPT, as illustrated in Figure \ref{ChatGPT}, incorporates the roles of the user, assistant (\ChatGPT), and system. First, a series of conversation histories between the user and the assistant are input as the prompts for \ChatGPT\, to ensure continuity in the conversation. Second, the system prompt instructs \ChatGPT\, to pretend it can view the images while reminding it to avoid discussing images in its response if the user's question is unrelated to them. Finally, the retrieved descriptions for the user's uploaded images are combined with the user's current question to form the user's prompt for \ChatGPT. Once \ChatGPT\, receives the prompt, it generates a response, which is then displayed by Boon.

\subsection{The Front-End of Boon}
\begin{figure}[htbp]
\centering
\begin{subfigure}[b]{0.7\linewidth}
    \includegraphics[width=\linewidth]{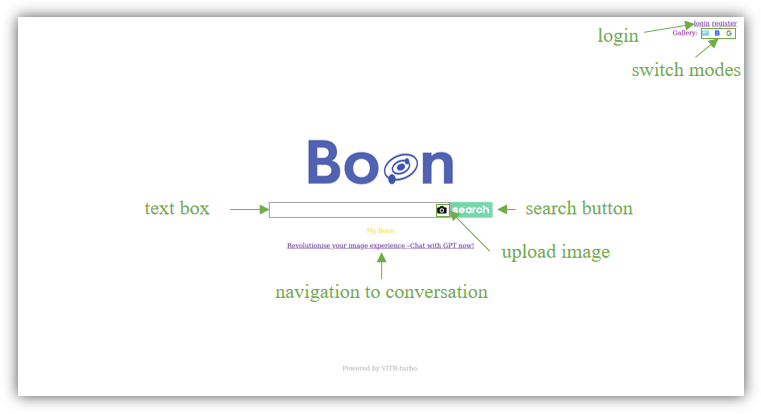}
    \caption{Navigation interface.}
    \label{fig:interface}
\end{subfigure}
\\ 
\begin{subfigure}[b]{0.7\linewidth}
    \includegraphics[width=\linewidth]{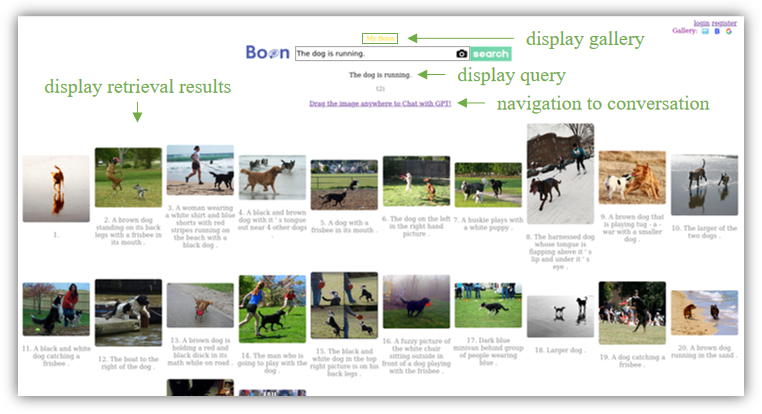}
    \caption{Retrieval interface.}
    \label{fig:retrieval}
\end{subfigure}
\\ 
\begin{subfigure}[b]{0.7\linewidth}
    \includegraphics[width=\linewidth]{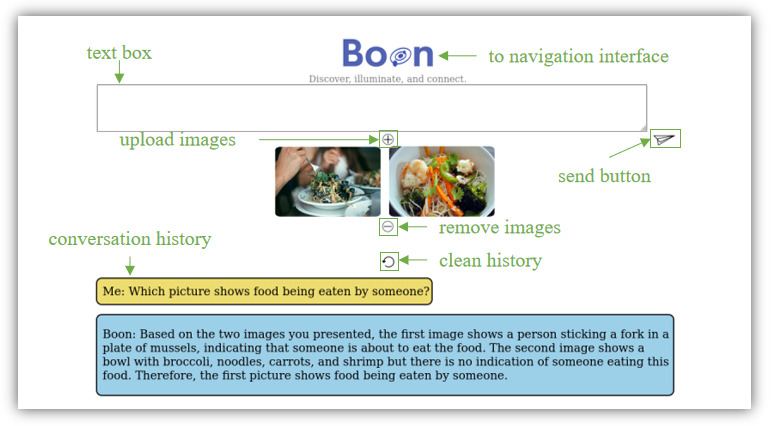}
    \caption{Conversation interface.}
    \label{fig:chat}
\end{subfigure}
\caption{The front-end components of Boon, include (a) the navigation interface, (b) the retrieval interface, and (c) the conversation interface.}
\label{ViewPages}
\end{figure}

Figure \ref{ViewPages} illustrates the front-end components of Boon, which include (a) the navigation interface, (b) the retrieval interface, and (c) the conversation interface. In the navigation interface, users can input a textual query in the provided text box or upload an image query using the upload button. Once the search button is clicked, the retrieval results are displayed on the retrieval interface. Both the navigation and retrieval interfaces include a button for navigating to the conversation interface. Within the conversation interface, users can input textual prompts in the text box and upload images using the upload button. After clicking the send button, the conversation history appears on the conversation interface.

To enhance the user experience, Boon incorporates mouse actions. \begin{itemize}
    \item When a user double-clicks a displayed description, the clicked description becomes the textual query for a text-to-image retrieval request. 
    \item When a user double-clicks a retrieved image, the clicked image becomes the image query for an image-to-text retrieval request. 
    \item When a user drags a retrieved image, the dragged image is transferred to the conversation interface as the prompt image.
\end{itemize}

\section{Results}

This section presents retrieval request results, including visuals of multi-lingual results, re-ranking for web images via Google's API, and quantitative findings. It also highlights Boon's image-related conversation requests with visuals and quantitative outcomes.

\subsection{Implementation Details}
A high-performance PC with a single NVIDIA RTX 3080 graphics card and 64GB of memory can meet the minimum requirements for running Boon. The proposed Boon was implemented using the Django framework. For the \VITR\, module, the model $\textrm{VITR}_{\textrm{L}}$ \cite{gong2023vitr} utilising the encoder of `ViT-L/14' from CLIP was employed, and the turbo setting $N$ was set to 200. The code for \VITR\, was implemented with the PyTorch framework. 

\subsection{Results of Retrieval Requests}

\begin{figure*}[htbp]
\centering
\begin{subfigure}[b]{1\linewidth}
    \includegraphics[width=\linewidth]{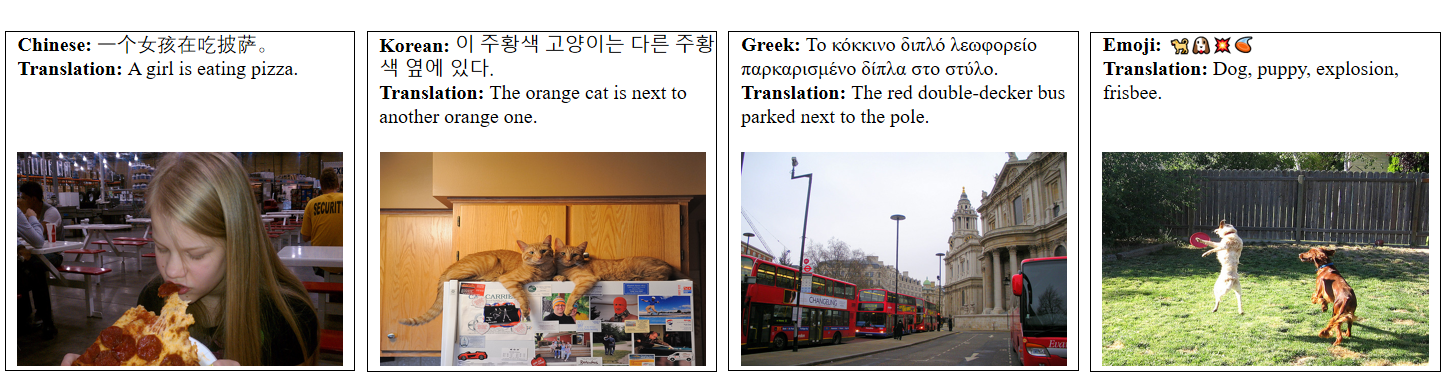}
    \caption{The top images retrieved using queries in Chinese, Korean, Greek, and Emoji languages.}
\end{subfigure}
\\  
\begin{subfigure}[b]{1\linewidth}
    \includegraphics[width=\linewidth]{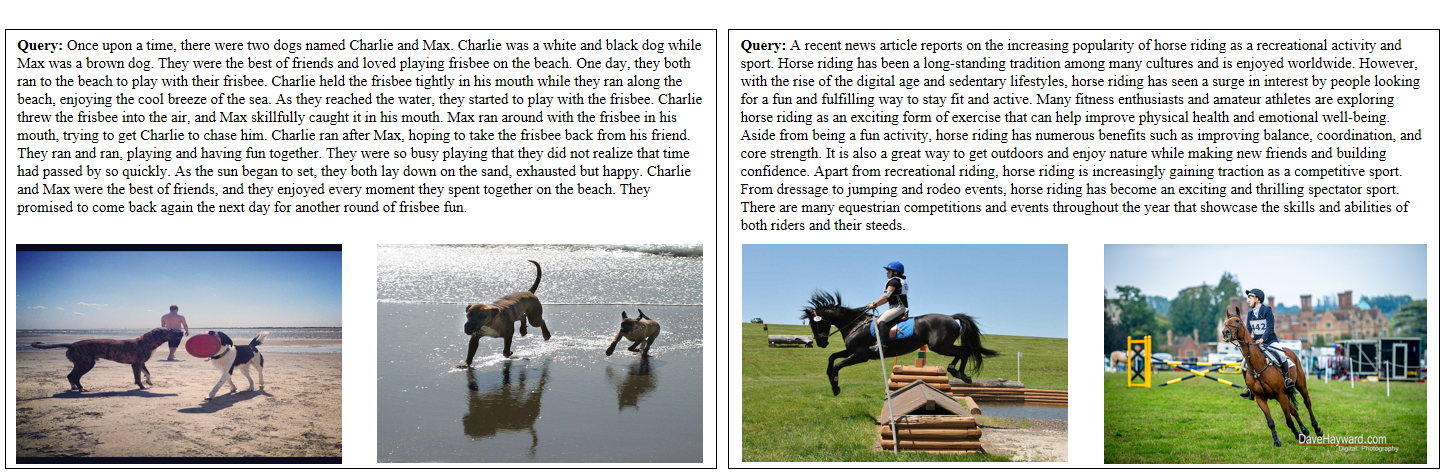}
    \caption{The top two images retrieved using long textual queries.}
\end{subfigure}
\\ 
\begin{subfigure}[b]{1\linewidth}
    \includegraphics[width=\linewidth]{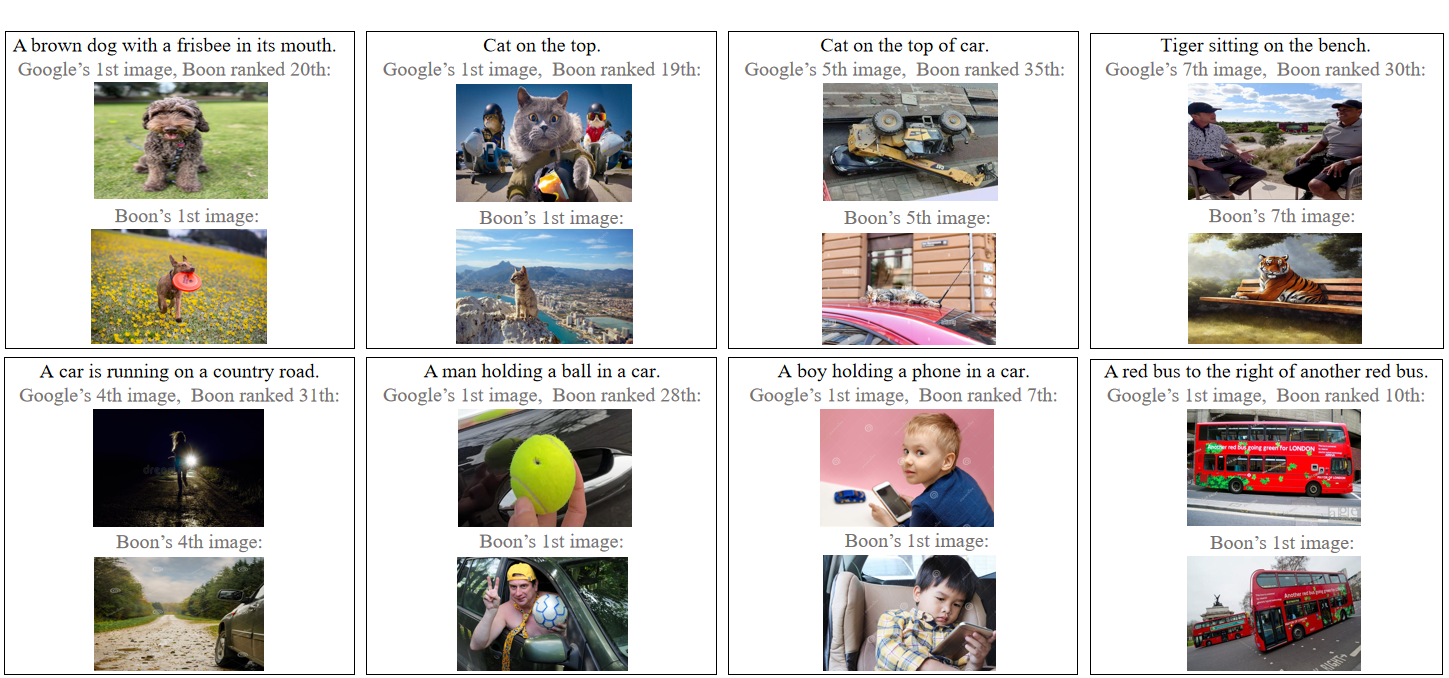}
    \caption{Compare the rankings of retrieved images in response to queries between Google's API and Boon, where Boon corrects and improves Google's erroneous results.}
\end{subfigure}
\caption{The retrieval examples for Boon include: (a) translating non-English queries for retrieval; (b) using long textual queries for retrieval; and (c) re-ranking images to the queries retrieved by Google's API.}
\label{RetrievalResults}
\end{figure*}

\textbf{Presenting Examples of Retrieval Requests.} Figure \ref{RetrievalResults} visually presents several retrieval results using non-English, English, and long textual queries, as well as the re-ranking results for images on the web retrieved through Google's API.

Figure \ref{RetrievalResults} (a) displays four examples of the top relevant result when employing non-English textual queries for retrieval. Four languages Chinese, Korean, Greek, and Emoji were tested. Each language was translated into English by the \ChatGPT\, module of Boon before being used by the \VITR\, module to search for relevant images. The \ChatGPT\, supported over \num{100} different languages for translation.

Figure \ref{RetrievalResults} (b) showcases two examples of the top two relevant results when using long textual queries (written in English) for retrieval. The first example's query was a story about two dogs, while the second example's query was a news article about horse riding. The results for both examples were relevant to their respective queries.

Figure \ref{RetrievalResults} (c) demonstrates examples of Boon re-ranking the retrieval results from Google's API. Considering that transferring images from Google's  API to Boon takes time, and users' focus is typically the top retrieved results, Boon obtains \num{40} retrieval results for each query using Google's API. It then recalculates the relevance between these \num{40} retrieved results and the query to re-rank them. 
In Figure \ref{RetrievalResults} (c), the top retrieved images by Google's API were irrelevant to the queries. Meanwhile, Boon re-ranked the retrieval results to position these images at lower rankings, and the top retrieved images by Boon were presented for comparison.
For example, in the second scenario, the user was searching for a picture of a cat on top of an object. However, the top retrieved image by Google's API featured the movie `Top Gun' with cats, ignoring the relation expressed in the user's query. Boon then re-ranked this image to the 19th position in terms of relevance and presented a more relevant image at the top.   

\textbf{Quantitative Results.} The retrieval performance of Boon was quantitatively evaluated using \VITR\, \cite{gong2023vitr}. Table \ref{BoonRefCOCOgResults} and Table \ref{BoonFlickr30KResults} compare the proposed Boon with baseline methods on the relation-focused dataset RefCOCOg \cite{mao2016generation} and the benchmark dataset Flickr30K \cite{young2014image}, respectively, for both image-to-text and text-to-image retrieval \cite{gong2023vitr}. The evaluation measure used is Recall at rank k (Recall@k), which is defined as the percentage of relevant items among the top k retrieved results \cite{saracevic1995evaluation}. In the RefCOCOg test set, as shown in Table \ref{BoonRefCOCOgResults}, Boon achieved average Recall@1 values of 45.2\% for image-to-text retrieval and 29.5\% for text-to-image retrieval, outperforming \gls{CLIP} by 2.8\% and 4.3\% respectively. In the Flickr30K test set, as shown in Table \ref{BoonFlickr30KResults}, Boon achieved average Recall@1 values of 94.7\% for image-to-text retrieval and 82.5\% for text-to-image retrieval, outperforming \gls{CLIP} by 2.1\% and 4.7\% respectively.

\begin{table}[htbp]
\caption{Results of cross-modal information retrieval networks on the RefCOCOg test set. Table shows average Recall@$k$ (\%) values.}
\label{BoonRefCOCOgResults}
\centering
\begin{tabular}{lcccccc}
\hline
\multirow{2}{*}{Network} & \multicolumn{3}{c}{Image-to-Text} & \multicolumn{3}{c}{Text-to-Image} \\
\cmidrule(lr){2-4} \cmidrule(lr){5-7} 
                         & R@1           & R@5           & R@10          & R@1           & R@5           & R@10          \\ \hline
VSRN++ \cite{gong2023vitr}                 & 20.0          & 44.9          & 57.3          & 13.8          & 34.6          & 47.8          \\
VSE$\infty$ \cite{gong2023vitr}            & 31.1          & 58.3            & 69.7          & 19.5          & 42.8         & 55.2           \\
CLIP \cite{gong2023vitr}                    & 42.4            & 65.5           & 75.1         & 25.2           & 48.9           & 60.4           \\
\hline
Boon               & \textbf{45.2} & \textbf{71.1} & \textbf{80.5} & \textbf{29.5} & \textbf{55.1} & \textbf{66.8} \\ \hline
\end{tabular}
\end{table}

\begin{table}[htbp]
\caption{Results of cross-modal information retrieval networks on the Flickr30K test set. Table shows average Recall@$k$ (\%) values.}
\label{BoonFlickr30KResults}
\centering
\begin{tabular}{lcccccc}
\hline
\multirow{2}{*}{Network} & \multicolumn{3}{c}{Image-to-Text} & \multicolumn{3}{c}{Text-to-Image} \\
\cmidrule(lr){2-4} \cmidrule(lr){5-7}
                         & R@1           & R@5           & R@10          & R@1           & R@5           & R@10          \\ \hline
VSRN++ \cite{li2022image}                  & 79.2          & 94.6          & 97.5          & 60.6          & 85.6          & 91.4          \\
VSE$\infty$ \cite{chen2021learning}            & 88.7          & 98.9          & 99.8          & 76.1          & 94.5          & 97.1          \\
CLIP \cite{gong2023vitr}                    & 92.6            & 99.2            & 99.6          & 77.8           & 95.2           & 97.7          \\ 
\hline
Boon               & \textbf{94.7} & \textbf{99.7} & \textbf{99.9} & \textbf{82.5} & \textbf{96.7} & \textbf{98.3} \\
\hline
\end{tabular}
\end{table}

\textbf{Assessing Retrieval Time for Each Query.}
The average retrieval time for each query using Boon was experimentally assessed. For image-to-text retrieval involving \num{21899} textual descriptions, the average retrieval time was 0.19 seconds. Meanwhile, for text-to-image retrieval encompassing \num{25799} images, the average retrieval time was 0.68 seconds.

\subsection{Results of Conversation Requests}

\begin{figure*}[htbp] 
\centering 
\includegraphics[width=1\linewidth]{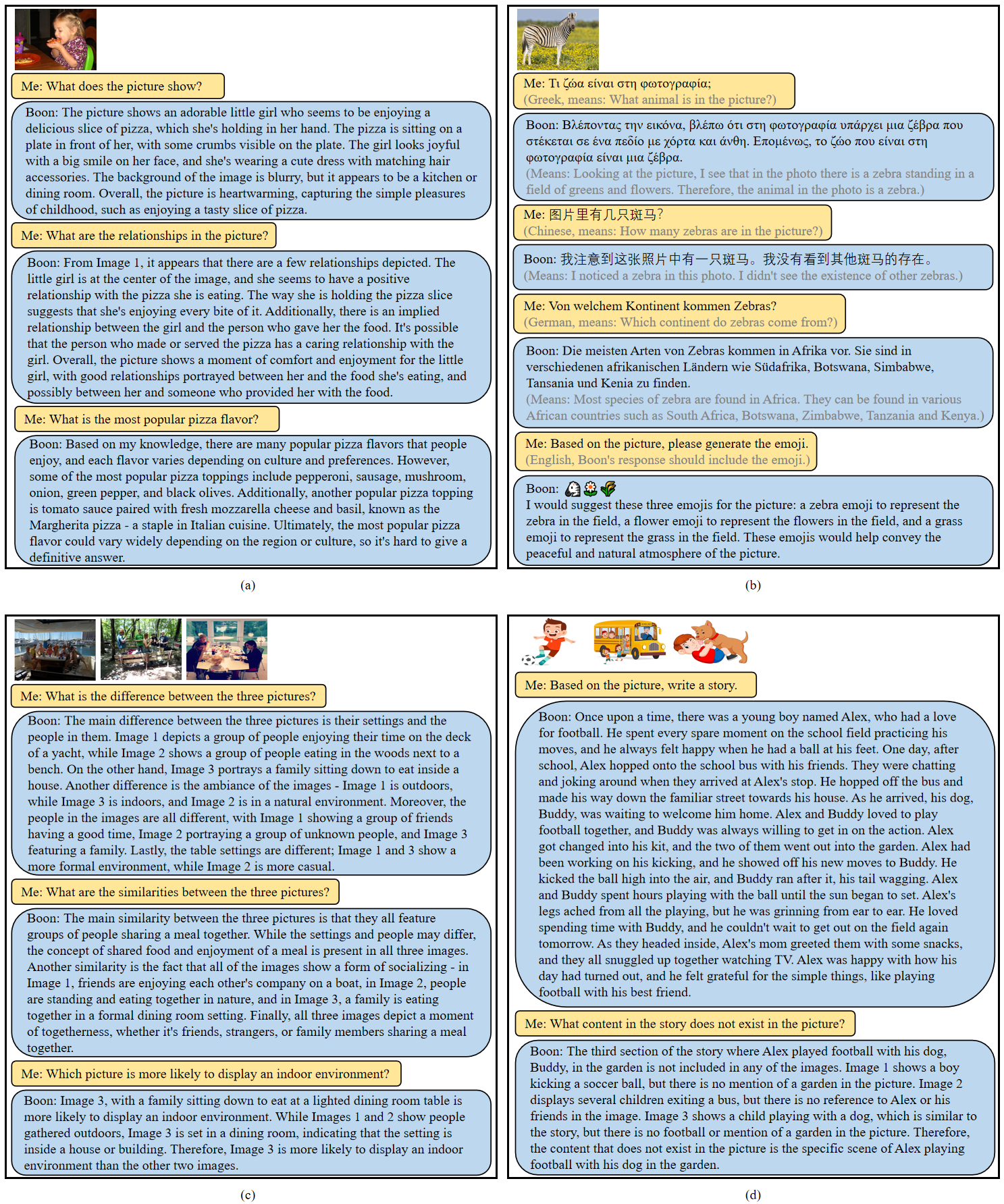} 
\caption{Examples of conversation requests about images include: (a) conversations about a single image, (b) multi-lingual conversations about an image, (c) conversations about multiple images, and (d) writing a story based on multiple images.} 
\label{ConversationResults} 
\end{figure*}

\textbf{Presenting Examples of Conversation Requests about Images.}
Figure \ref{ConversationResults} showcases four examples of conversation requests about images in Boon. 
In Figure \ref{ConversationResults} (a), a series of conversations revolve around a picture of a girl eating pizza, such as those regarding relationships in the picture. 
In Figure \ref{ConversationResults} (b), Boon showcases its multi-lingual proficiency in conversation about images. A zebra picture was uploaded, and questions were asked in Greek, Chinese, German, and English. Boon accurately responded in the respective languages and generated the appropriate emoji as the response based on the image as requested.
Figure \ref{ConversationResults} (c) demonstrates that Boon can support multiple prompt images and answer questions about their differences and similarities. 
In Figure \ref{ConversationResults} (d), Boon has written a story based on three related pictures. The story is about a boy who played football at school and then played with his dog after school. Unlike other visual question-answering networks \cite{shih2016look, yu2020cross}, Boon can continuously communicate with users and answer high-level semantic questions.

\textbf{Quantitative Results.} Ensuring that the retrieved descriptions by Boon accurately describe the prompt images is the key to fulfilling the conversation request regarding images. To evaluate this, the paper utilises the Flickr30K dataset \cite{young2014image}, and the descriptions in the dataset were not included in Boon's description pool. Specifically, Boon retrieved relevant descriptions from its description pool for the \num{1000} images in the Flickr30K test set and compared them with the images' corresponding descriptions in the dataset. BertScore, which utilises contextual embeddings from BERT to compare the similarity between two pieces of text \cite{zhang2019bertscore}, was used as an evaluation measure  with a maximum value of 1. 
According to the evaluation results, Boon's average BertScore on the Flickr30K test set was 0.91.

\section{Conclusion}
VSE networks improve search engine accuracy by associating visual content with relevant text. They can be used in cross-modal search engines to retrieve multimedia resources, by embedding image and textual descriptions in a shared latent space. This paper introduces a novel cross-modal search engine, Boon, which improves user experience in image-to-text and text-to-image retrieval tasks by incorporating the cutting-edge \gls{VSE} network, VIsion Transformers with Relation-focused learning (\gls{VITR}), and ChatGPT that is an advanced Large Language Model (LLM). 
Boon leverages \gls{VITR} to emphasise information retrieval based on user query relations and enhance both image-to-text and text-to-image retrieval performance. 
Furthermore, it utilises ChatGPT to facilitate translations in multiple languages and enable conversations about images, broadening accessibility for various audiences, including visually impaired individuals. 
By supplying relevant image descriptions obtained from Boon's integrated \gls{VITR} as input prompts, the limitations of ChatGPT's 3.5 model in comprehending images are overcome. 
The interface and capabilities of Boon's search engine are demonstrated using the RefCOCOg dataset, and its ability to search for multimedia online is facilitated by Google's API. 
Future developments for Boon will involve implementing text-video retrieval functions. Furthermore, a deep comparison between Boon and existing search engines like Google and Bing will be part of future work.

\bibliographystyle{unsrt}  
\bibliography{egbib}  

\end{document}